# Data Transmission with Reduced Delay for Distributed Acoustic Sensors

Hyun-Gyu Ryu, Sang-Keum Lee, and Dongsoo Har

Cho Chun Shik Graduate School for Green Transportation at

Korea Advanced Institute of Science and Technology, Daejeon, Korea

e-mail: dshar@kaist.ac.kr

Abstract

This paper proposes a channel access control scheme fit to dense acoustic sensor nodes in a sensor network. In the considered scenario, multiple acoustic sensor nodes within communication range of a cluster head are grouped into clusters. Acoustic sensor nodes in a cluster detect acoustic signals and convert them into electric signals (packets). Detection by acoustic sensors can be executed periodically or randomly and random detection by acoustic sensors is event driven. As a result, each acoustic sensor generates their packets (50bytes each) periodically or randomly over short time intervals (400ms~4seconds) and transmits directly to a cluster head (coordinator node). Our approach proposes to use a slotted carrier sense multiple access. All acoustic sensor nodes in a cluster are allocated to time slots and the number of allocated sensor nodes to each time slot is uniform. All sensor nodes allocated to a time slot listen for packet transmission from the beginning of the time slot for a duration proportional to their priority. The first node that detect the channel to be free for its whole window is allowed to transmit. The order of packet transmissions with the acoustic sensor nodes in the time slot is autonomously adjusted according to the history of packet transmissions in the time slot. In simulations, performances of the proposed scheme are demonstrated by the comparisons with other low rate wireless channel access schemes.

## I. Introduction

Acoustic sensor networks have been studied for a variety of applications such as hearing aids [1, 2] and acoustic monitoring [3, 4]. Each acoustic sensor node can detect activities of objects or human beings based on acoustic information [5]. Due to distribution of acoustic sensor nodes in a sensor network, efficient in-network information fusion is indispensable [6, 7]. The most typical approach to achieve in-network information fusion with distributed acoustic sensor nodes is to build a hierarchy and the nodes that form a cluster transmit their acoustic (microphone) signal(s) to a higher level node, referred to as a cluster head. Fig. 1 shows the information fusion with acoustic sensor nodes in a cluster and Fig. 2 presents hierarchical sensor network. In case of hierarchical sensor network, multiple clusters are formed and each cluster carries out distributed in-network information fusion. However, with large number of acoustic sensor nodes within a cluster, information fusion process might lead to a combinatorial problem [8, 9]. In order to mitigate the combinatorial problem, [10] address a potential solution. According to [10], each node enhances its own local microphone signal in an optimal way, as if all signals in the entire acoustic sensor network were available to each node. This type of algorithms are referred to as distributed adaptive node-specific signal estimation algorithms. Unfortunately, the optimality of the algorithms relies on the assumption that the total number of desired speakers is much smaller than the number of available microphones.

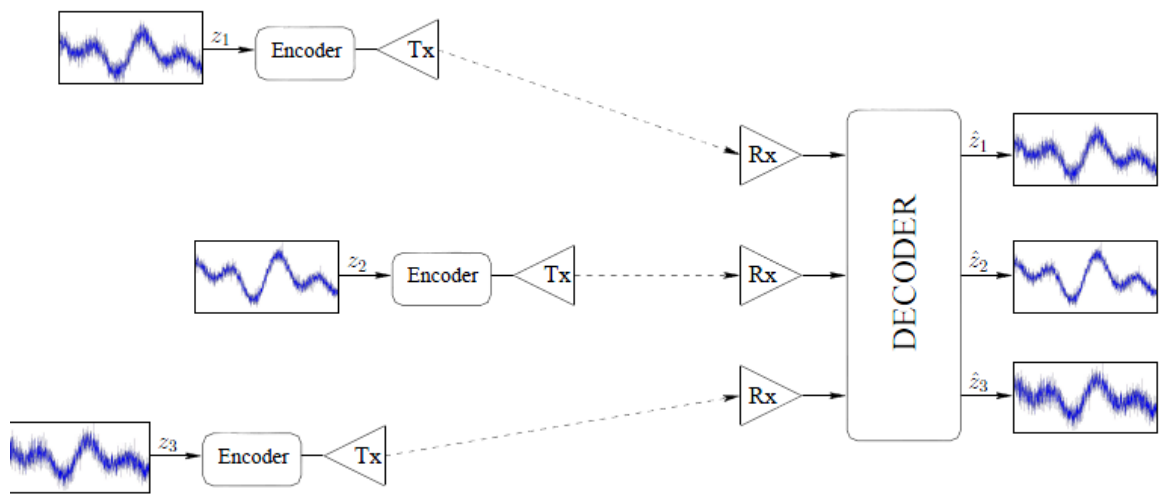

Fig. 1. Typical source coding in a cluster of acoustic sensor nodes : three nodes encode their locally preprocessed signals and transmit it to a fourth node, which decodes all three signals [11].

The acoustic signal is typically converted into electric signal and the digitized electric signal is wirelessly transmitted to the cluster head in the form of electromagnetic propagation. Hence, spectrum sensing for transmission scheduling is required. Deng et al. [12] devised sensor scheduling by grouping the sensors into non-disjoint subsets. A sensor network consisting of clusters with a hierarchical routing protocol in order to increase network lifetime was reported by Huang et al. [13]. They showed, with many sensor nodes, reduction of energy consumption by means of hierarchical routing instead of flat routing.

For efficient operation of networked sensor nodes over certain area, various channel access schemes for different types of network topologies have been studied [14, 15]. Particularly, for low rate data transmission, the channel access schemes such as the IEEE 802.15.4 standard (ZigBee) [16] and the BMAC [17] have been investigated. However, most of the previous channel access schemes require signaling overhead for each channel setup and their performances often depend on traffic conditions. In this paper, an efficient scheme that requires no signaling overhead and works comparatively well with different type of traffic is presented.

Organization of this paper is as follows. Section II addresses cluster formation and channel access control schemes. Section III gives simulation results to validate the effectiveness of the proposed channel access scheme and Section IV concludes this paper.

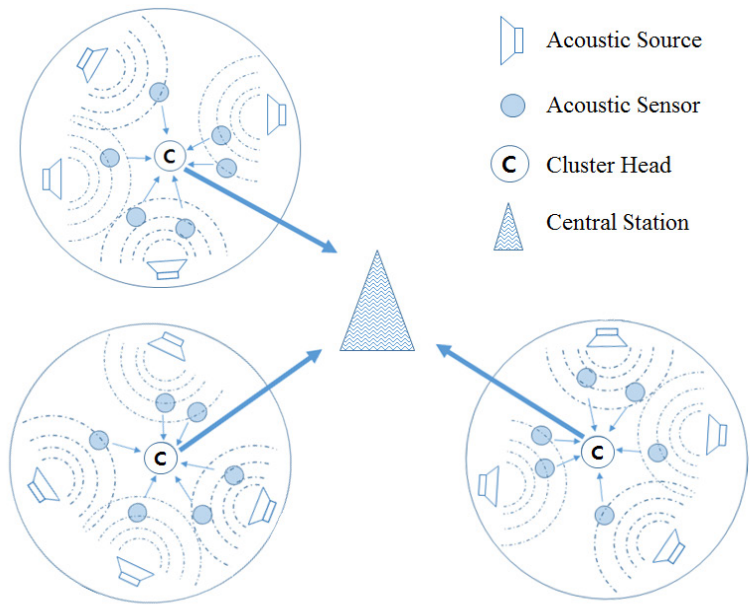

Fig. 2. Clustered acoustic sensor nodes located around the cluster heads.

## II. System Description

### A. Cluster Formation

Consider the acoustic sensor nodes located around the cluster heads in Fig. 2. The location of each acoustic sensor node is assumed fixed. It is assumed that the cluster head and the sensor nodes in a cluster operate in a time-slotted fashion for timely communication. Depending on geometrical distance between neighbored clusters, frequency reuse pattern among clusters is determined. The acoustic sensor nodes send (report) the sensing results directly to the cluster head and further forwarded to the central station.

For initial clustering purpose, cluster head broadcasts an registration (RGT) message which contains the identification number (ID) of the cluster head, its position, and a header field. The purpose of the header field is to differentiate the advertisement message from other types of message or data. The format of the RGT message is given as follows

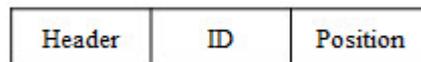

Acoustic sensor nodes within communication range from the cluster head respond by sending a join request (J_REQ), which consists of the identification number of the sensor node (N_ID), the signal-to-noise ratio (SNR) of the received RGT message, and the identification number of the destination cluster head (CH_ID). The format of the J_REQ is

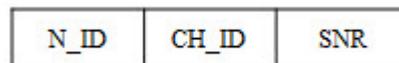

An acoustic sensor node may receive multiple RGT messages from different cluster heads. In this case, the sensor node will join the cluster head that is closest to it in order to consume the minimum transmission energy. Notice that a sensor node knows the position(s) of the cluster head(s) via the RGT messages. Flow chart of clustering process can be shown in Fig.3.

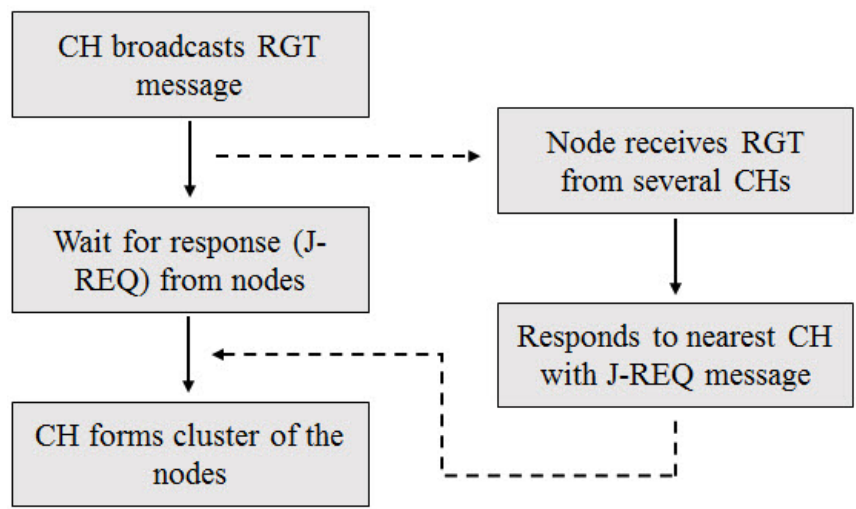

Fig. 3. Flowchart of clustering process.

**B. Determination of cluster size**

It is crucial to determine the number of sensor nodes in a cluster, which ensures the minimum level of sensing performance. Sensing performance can be defined in terms of minimum global detection probability and maximum false alarm probability.

The detection probability of a sensor node is defined as the probability that a sensor node correctly detects the presence of the desired acoustic signal. On the other hand, false alarm probability is defined as the probability that a sensor node incorrectly detects the presence of the desired acoustic signal when the acoustic signal is actually absent. The detection probability $P_{dj}$ and the false alarm probability $P_{fj}$ of the $j$-th sensor node of a cluster can be given as follows [18]

$$P_{dj} = Q_u\left(\sqrt{2\gamma_j}, \sqrt{\varepsilon}\right) \quad (1a)$$

$$P_{fj} = \frac{\Gamma\left(u, \frac{\varepsilon}{2}\right)}{\Gamma(u)} \quad (1b)$$

where $\gamma_j$ is the SNR at the $j$-th node and $\varepsilon$ denotes the energy threshold for a local decision and $u$ represents the number of samples and $\Gamma(.,.)$ is the incomplete gamma function and $\Gamma(.)$ is the complete gamma function and $Q_u(.,.)$ is the generalized Marcum Q-function. Note that $\gamma_j$ is reported as SNR to the cluster head as a part of the J_REQ message. The decision fusion at the cluster head often employs the OR-rule, which decides the presence of the desired acoustic signal when at least one of the acoustic sensor nodes reports its presence. Let $S$ be the number of sensor nodes in a cluster, then the global detection probability $Q_d$ and the global false alarm probability $Q_f$ achieved by $S$ sensor nodes in a cluster, employing the OR-rule, are given, respectively, as

$$Q_d = 1 - \prod_{j=1}^{S}(1 - P_{dj}) \quad (2a)$$

$$Q_f = 1 - \prod_{j=1}^{S}(1 - P_{fj}). \quad (2b)$$

and $Q_d$ and $Q_f$ must satisfy the required performance level as follows:

$$Q_d \geq Q_d^{min} \tag{3a}$$

$$Q_f \leq Q_f^{max} \tag{3b}$$

where $Q_d^{min}$ is the minimum global detection probability required and $Q_f^{max}$ is the maximum global false alarm probability allowed. Due to conversion to electric signal, foregoing framework is identical with typical spectrum sensing [19]. Based on the derivation result in [20], the $S$ can be obtained as

$$S = \left\lfloor \frac{\log(1-Q_f^{max})}{\log(1-P_f^{max})} \right\rfloor \tag{4}$$

**C. Channel Access Control for Acoustic Sensor Nodes**

Wireless acoustic sensor networks are desired to have energy efficiency, low latency, high throughput, and fairness. In case of low rate wireless sensor networks, many works have been published so far. Among them, IEEE 802.15.4 (ZigBee) and BMAC are popularly adopted due to their attributes well matched with those required for wireless sensor networks.

**1) IEEE 802.15.4**

IEEE 802.15.4 standard specifies the medium access control (MAC) sub-layer/(physical)PHY layer of ZigBee [16]. IEEE 802.15.4 standard employs time division multiple access (TDMA) and carrier sense multiple access/collision avoidance(CSMA/CA). IEEE 802.15.4 complying system can be operated in beacon mode and non-beacon mode. In beacon mode, sensor nodes are synchronized by beacon signal which is transmitted by coordinator node, e.g., cluster head. Superframe interval is directly defined by beacon interval. The beacon interval is categorized into active portion and inactive portion. In the inactive portion, power consumption is minimized by turning off the cluster head and the transceiver of each sensor node. The active portion is divided into contention access period (CAP) and contention free period (CFP). The number of time slots in active portion is limited up to 15. Maximum number of time slots in CFP is limited to 7. In CAP, each node transmits information by CSMA/CA method where nodes access the common channel by carrier sensing. In CFP, the cluster head arranges the order of transmission and allocates number of time slots for sensor nodes. While IEEE 802.15.4 standard leads to high energy efficiency, most sensor nodes have to compete for channel access due to limited number of time slots for CFP.

## 2) BERKELEY MAC (BMAC)

BMAC is a contention based MAC protocol which is widely used in sensor networks. The BMAC is like the Aloha protocol with preamble sampling and the BMAC duty cycle determines the operation pattern of the radio transceiver [21]. The preamble length is provided as a parameter to the upper layer and it determines an optimal trade-off between energy savings and latency. Fig. 4 shows the preamble sampling of the BMAC. The BMAC is also similar to CSMA from the perspective of low power consumption. Unsynchronized duty cycling and long preambles are used in the BMAC to wake up receivers. Filter mechanism of the BMAC increases reliability and channel assessment. The sensor node operating with the BMAC can change operating variables such as back off values. The BMAC also adopts an adaptive preamble sampling scheme which can minimize idle listening and reduce duty cycle. The clear channel assessment (CCA) technique is used by the BMAC to decide if a packet is arriving when the sensor node wakes up. If no packet arrived, timeout puts node back to sleep. The CCA and packet bakeoffs are used by the BMAC for channel arbitration and link layer acknowledgments for reliability. There are no synchronization, RTS, and CTS adopted in BMAC.

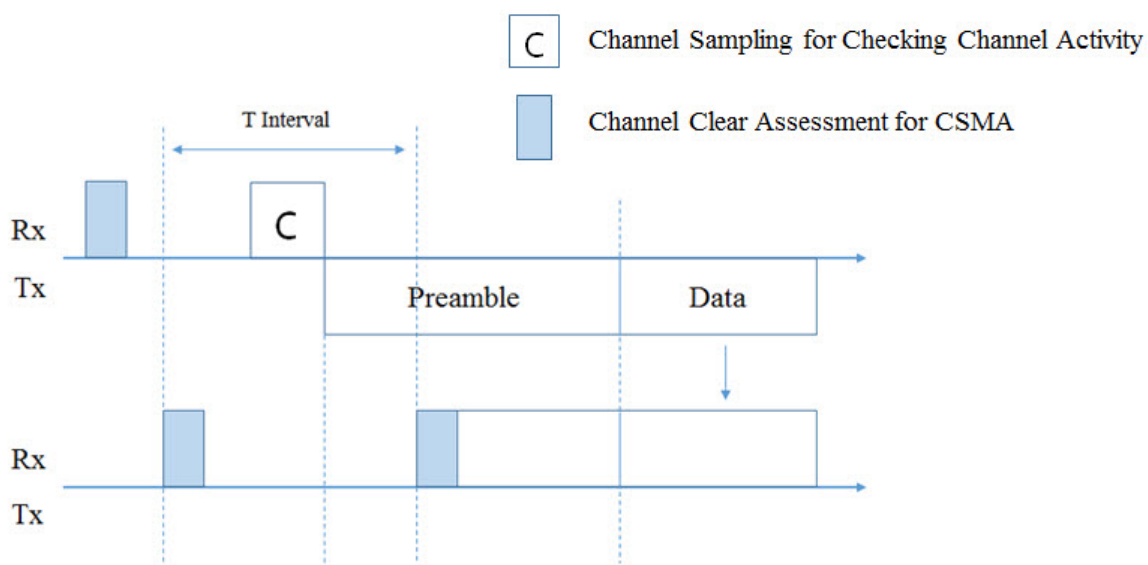

Fig. 4. Preamble sampling in BMAC.

## 3) Proposed Channel Control Scheme

Proposed channel control scheme achieves time synchronization between sensor nodes and cluster head by the beacon signal which is also used for such purpose by IEEE 802.15.4 complying systems. Sensor nodes are uniformly allocated to each time slot. One feature of the proposed channel control scheme is that queueing of sensor nodes allocated to each time slot is managed by sensor nodes themselves rather than the cluster head.

Consider the uplink packet transmission in Fig. 5(a). N nodes are allocated to each time slot and 15 time slots are in each repeated superframe. In a time slot, (N-1) windows in Fig. 5(b) for carrier sensing are co-located. Purpose of this in-slot window is to sense the carrier signal in shared channel. Minimum duration of the window ($W_1$) is set to the time resolution determined by the system and the durations of the other windows satisfy $W_k = k \times W_1$, k=2,...,N-1. Packets of the N nodes for transmission over the time slot can be modeled by the packets in a single virtual queue waiting to be served by a server (Fig.5(c)). Only one packet is transmitted at a time for each node, so the position of the packet corresponds to the position of the associated node. Initial order of packet transmission across the N nodes is preset by the system. Once the initial order is set, no additional coordination is needed for future packet transmissions by the N nodes.

Queue operation of the proposed scheduling in Fig. 5(c) can be explained in conjunction with Fig. 5(b). In Fig.5(c) with 5 nodes, the 3rd node is in the Head Of Line (HOL) position for immediate transmission. If it has a packet to transmit from the beginning of the time slot, it transmits the packet and takes the last position of the queue. Positions of the other nodes in the queue are shifted by one toward the HOL position. If the 3rd node has no packet to transmit, it remains at the same position and the 4th node that sensed the channel over the time duration of $W_1$ has a chance for transmission. If it has a packet to transmit, it does so immediately after $W_1$ expires and its position in the queue becomes the last. If the 4th node transmits its packet, the queue position of the 3rd node is unchanged and the queue positions of the other nodes behind the 4th node are shifted by one to the HOL position. When both the 3rd and the 4th nodes have no packets to transmit, i.e., their positions in the queue are not changed, the 5th node that senses the channel over the time duration $W_2$ has a chance for packet transmission after $W_2$ expires. With its transmission, queue positions of the other nodes after it are shifted by one toward the HOL position. If no nodes have a packet to transmit before the beginning of the time slot, positions of all the nodes are kept intact. As a whole, each node knows its position in the queue by the sensing time and the number of packet transmissions.

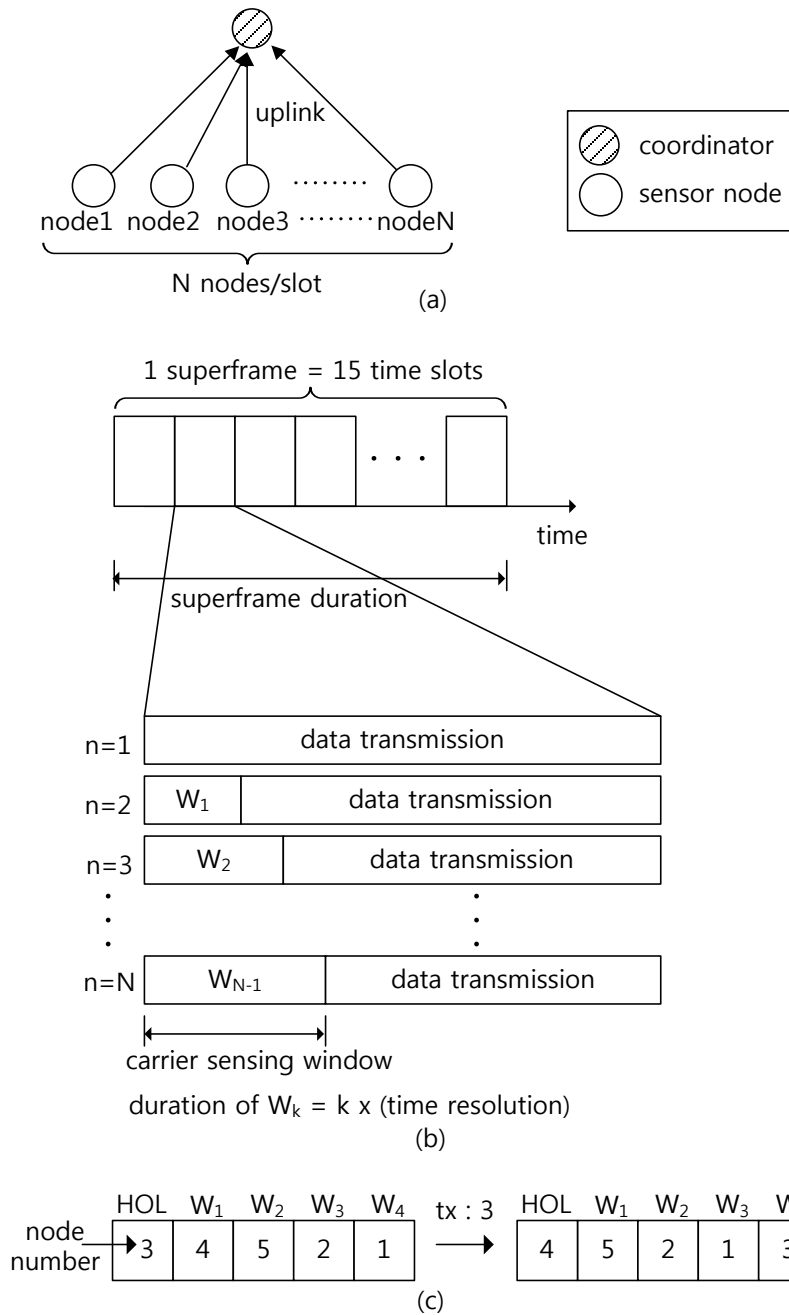

Fig. 5. Proposed channel control scheme for the acoustic sensor nodes in a cluster : (a) System model; (b) carrier sensing windows; (c) queue operation. Coordinator corresponds to cluster head.

## III. Simulations for Performance Comparisons

In this section, simulation results of the proposed scheme, IEEE 802.15.4, and the BMAC are presented. Comparisons are made in terms of average delay per packet and average energy consumption per packet.

Table 1. Common simulation parameters.

| Parameter | Value |
|---|---|
| packet generation interval (fixed interval generation) | 400/800/1200/1600 ms |
| packet generation interval (random generation) | 1000/2000/3000/4000 ms |
| data transmission rate | 100 kbps |
| packet size | 50 byte |
| tx/idle power | 35/41 mW |

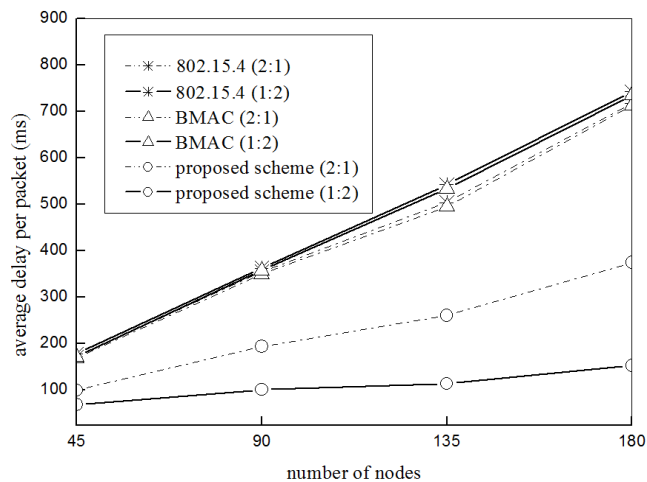

(a)

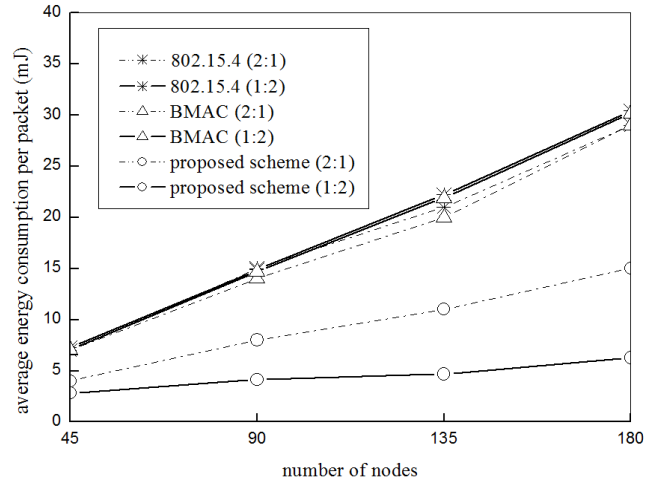

(b)

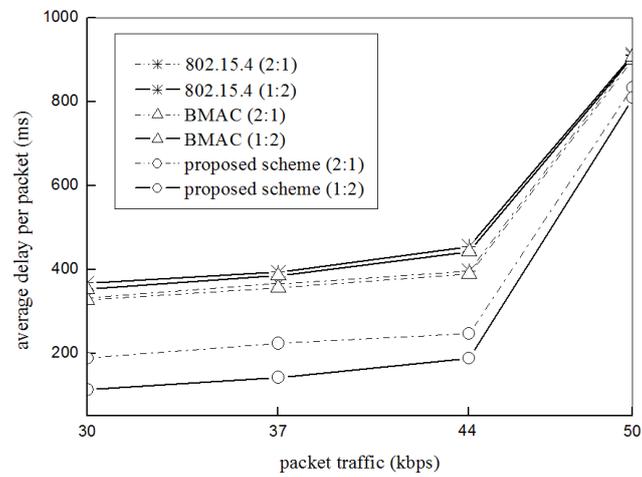

(c)

Fig. 6. Comparison of performance measures : (a) average delay per packet versus number of nodes per time slot; (b) energy consumption per packet versus number of nodes per time slot; (c) average delay per packet versus packet traffic.

Simulation parameters are shown in Table 1. Number of time slots and the superframe duration for IEEE 802.15.4 and the proposed scheme are 15 and 120ms, respectively. For the purpose of comparison, 15 slots for the IEEE 802.15.4 are dedicated to contention-free access. The smallest window duration $W_1$ is set to 250us. For the BMAC, no time slot is considered. Also for comparison, no signaling overhead is taken into account even for the IEEE 802.15.4 and the BMAC. Fixed packet generation intervals adopted for simulations are 400ms, 800ms, 1200ms, 1600ms. Random packet generation intervals used for simulations are 1000ms, 2000ms, 3000ms, 4000ms. Data transmission rate of the network is 100kbps. Packet size is 50 bytes.

Figure 6 shows average delay per packet and average energy consumption per packet. In case of the IEEE 802.15.4 and the proposed scheme, total nodes pertinent to the horizontal axes of Fig. 6 are equally divided over 15 slots. Delay and energy consumption are measured from the generation to the end of transmission. Aggregate packet traffic of the nodes is adjusted so that channel utilization is less than 1. Numbers on horizontal axes in Fig.6 match with the numbers in Table 1. For example, the 2:1 with 9 nodes per slot (9 nodes/slot * 15 slots=135 nodes = 3rd entry on axis) indicates that 6 nodes generate packets in fixed interval 1200ms (3rd entry of fixed interval) and 3 nodes generate packets randomly over 3000ms interval (3rd entry of random generation). When it is 2:1 data traffic, performances of the considered schemes allow us to predict the performances of all the schemes with purely periodic traffic. Also with 1:2 data traffic, the performances of the considered schemes with purely random traffic can be estimated. As seen in Fig.6, the proposed scheme with autonomous queue operation is superior to the IEEE 802.15.4 based scheme and the BMAC scheme. Disparity of performance is getting more outstanding as the number of nodes increases. Average energy consumption in Fig. 6(b) seems proportional to average packet delay in Fig. 6(a). Interestingly, the proposed scheme works significantly better with random traffic dominant in aggregate traffic, whereas no such differentiation between type of packet traffic is observed with the IEEE 802.15.4 based scheme and the BMAC scheme. From the simulation results, carrier sensing in in-slot windows is effective for reduction of average packet delay and average energy consumption.

It is noted that the proposed scheme can be extended for hierarchical wireless network with multiple levels. A mother node in upper level of hierarchy having multiple child nodes can take the role of coordinator node and each child node in lower level attempts to transmit data following the order of data transmission.

## IV. Conclusion

This paper proposes a channel access control scheme fit to dense acoustic sensor nodes in a sensor network. Multiple acoustic sensor nodes are grouped into clusters and the acoustic sensor nodes of each cluster transmit detected information to the cluster head.

Detection by acoustic sensors can be executed periodically or randomly and random detection by acoustic sensors is event driven. Our approach based on a slotted carrier sense multiple access. All acoustic sensor nodes allocated to a time slot listen for packet transmission from the beginning of each slot for a duration proportional to their priority, and transmit detected information to the cluster head. The order of packet transmissions with the acoustic sensor nodes in the time slot is autonomously adjusted according to the history of packet transmissions in the time slot. By simulations, superior performances of the proposed scheme in terms of transmission delay and average energy consumption are demonstrated by the comparisons with other low rate wireless channel access schemes.

## Acknowledgement

This research was supported by Ministry of Land, Infrastructure, and Transport (MOLIT) as Railroad Specialized Graduate School.

## References

[1] B. Cornelis, S. Doclo, T. Van dan Bogaert, M. Moonen, and J. Wouters, "Theoretical analysis of binaural multi microphone noise reduction techniques," IEEE Transactions on Audio, Speech, and Language Processing, vol. 18, no. 2, pp. 342–355, 2010.

[2] S. Doclo, T. van den Bogaert, M. Moonen, and J. Wouters, "Reduced bandwidth and distributed MWF-based noise reduction algorithms for binaural hearing aids," IEEE Transactions on Audio, Speech, and Language Processing, vol. 17, no. 1, pp. 38–51, 2009.

[3] H. Wang and P. Chu, "Voice source localization for automatic camera pointing system in videoconferencing," In Acoustics, Speech, and Signal Processing, IEEE International Conference on, vol. 1, pp. 187-190, 1997.

[4] Y. Guo and M. Hazas, "Acoustic source localization of everyday sounds using wireless sensor networks," Proceedings of the 12th ACM international conference adjunct papers on Ubiquitous computing, pp. 411–412, 2010.

[5] J. M. Sim, Y. Lee, and O. Kwon, "Acoustic Sensor Based Recognition of Human Activity in Everyday Life for Smart Home Services," International Journal of Distributed Sensor Networks, vol. 2015, Article ID 679123, 11 pages, 2015.

[6] G. Han, C. Zhang, L. Shu, N. Sun, and Q. Li, "A Survey on Deployment Algorithms in Underwater Acoustic Sensor Networks," International Journal of Distributed Sensor Networks, vol. 2013, Article ID 314049, 11 pages, 2013.

[7] N. Javaid, M. R. Jafri, S. Ahmed, M. Jamil, Z. A. Khan, U. Qasim and S. S. Al-Saleh, "Delay-Sensitive Routing Schemes for Underwater Acoustic Sensor Networks," International Journal of Distributed Sensor Networks, vol. 2015, Article ID 532676, 13 pages, 2015.

[8] Y. Jia, Y. Luo, Y. Lin, and I. Kozintsev, "Distributed microphone arrays for digital home and office," In Acoustics, Speech and Signal Processing, IEEE International Conference on, vol. 5, pp. V-V, 2006.

[9] I. Himawan, I. McCowan, and S. Sridharan, "Clustered blind beamforming from ad-hoc microphone arrays," IEEE Transactions on Audio, Speech, and Language Processing, vol. 19, no. 4, pp. 661–676, 2011.

[10] A. Bertrand and M. Moonen, "Distributed adaptive node-specific signal estimation in fully connected sensor networks – part I: sequential node updating," IEEE Transactions on Signal Processing, vol. 58, no. 10, pp. 5277–5291, 2010.

[11] A. Bertrand, "Applications and Trends in Wireless Acoustic Sensor Networks: a Signal Processing Perspective," Proceedings of 18th IEEE Symposium on Communications and Vehicular Technology in the Benelux (SCVT), pp. 1-6, 2011.

[12] R. Deng, J. Chen, C. Yuen, P. Cheng, and Y. Sun, "Energy-efficient cooperative spectrum sensing by optimal scheduling in sensor-aided cognitive radio networks," IEEE Transactions on Vehicular Technology, vol. 61, pp. 716-725, 2012.

[13] Z. Huang, Y. Cheng, and W. Liu, "A novel energy-efficient routing algorithm in multi-sink wireless sensor networks," IEEE 10th International Conference on Trust, Security and Privacy in Computing and Communications (TrustCom), pp. 1646-1651, 2011.

[14] C. Li, H. Li, and R. Kohno, "Performance evaluation of IEEE 802.15.4 for wireless body area network(WBAN)," Proceedings of IEEE International Conference on Communications Workshops, pp. 1-5, 2009.

[15] F. Subhan, H. Hasbullah, A. Rozyyev, and S. T. Bakhsh, "Analysis of bluetooth signal parameters for indoor positioning systems," In Computer & Information Science (ICCIS), International Conference on, vol. 2, pp. 784-789, 2012.

[16] A. Koubaa, M. Alves, and E. Tovar, "i-GAME: an implicit GTS allocation mechanism in IEEE 802.15. 4 for time-sensitive wireless sensor networks," IEEE 18th Euromicro Conference on Real-Time Systems, pp. 182-192, 2006

[17] J. Polastre, J. Hill, and D. Culler. "Versatile low power media access for wireless sensor networks," Proceedings of the 2nd international conference on Embedded networked sensor systems, pp. 95-107, 2004.

[18] T. Peng, C. Guo, and W. B. Wang, "Energy-efficient cooperative spectrum sensing in cognitive radio networks," Journal of Beijing University of Posts and Telecommunications, vol. 33, no. 4, pp. 93-96, 2010.

[19] E. Hong, K. Kim, and D. Har, "Spectrum sensing by parallel pairs of cross-correlators and comb filters for OFDM systems with pilot tones," IEEE Sensors Journal, vol. 12, no. 7, pp. 2380-2383, 2012.

[20] H. A. Shah, M. Usman, and I. Koo, "Bioinformatics-Inspired Quantized Hard Combination based Abnormality Detection for Cooperative Spectrum Sensing in Cognitive Radio Networks," IEEE Sensors Journal, vol. 15, no. 4, pp. 2324-2334, 2015.

[21] T. V. Dam and K. Langendoen, "An Adaptive Energy-Efficient MAC Protocol for Wireless Sensor Networks", In Proceedings of the 1st international conference on Embedded networked sensor systems, pp. 171-180, 2003.